\documentstyle[preprint,aps,eqsecnum]{revtex}

\def\beq#1{\begin{equation} \label{#1}}
\def\eeq{\end{equation}}
\def\bra#1{\left\langle #1\right\vert}
\def\ket#1{\left\vert #1\right\rangle}

\input prepictex
\input pictex
\input postpictex
\newdimen\tdim
\tdim=\unitlength
\def\stpltsmbl{\setplotsymbol ({\small .})}

\newbox\phru
\setbox\phru=\hbox{\beginpicture
\setcoordinatesystem units <\tdim,\tdim>
\stpltsmbl
\setquadratic
\plot
0 0
2.5 3
5 0
7.5 -3
10 0
/
\endpicture}
\def\photonru #1 #2 *#3 /{\multiput {\copy\phru}  at
#1 #2 *#3 10 0 /}

\newbox\sru
\setbox\sru=\hbox{\beginpicture
\setcoordinatesystem units <\tdim,\tdim>
\stpltsmbl
\setquadratic
\plot
  0.0   0.0
  4.8   1.5
  7.5   5.0
  7.3   8.5
  5.0  10.0
  2.7   8.5
  2.5   5.0
  5.2   1.5
 10.0   0.0
/
\endpicture}
\def\springru #1 #2 *#3 /{\multiput {\copy\sru}  at
#1 #2 *#3 10 0 /}

\def\beq#1{\begin{equation} \label{#1}}
\def\eeq{\end{equation}}
\def\bra#1{\left\langle #1\right\vert}
\def\ket#1{\left\vert #1\right\rangle}

\begin{document}
{
\tighten
\begin{center}
{\Large\bf
New $B^{\pm}\rightarrow K\pi$ data
explain absence of CP violation
Tree-penguin interference canceled by Pauli effects}
\vrule height 2.5ex depth 0ex width 0pt
\vskip0.8cm
Harry J. Lipkin\,$^{b,c}$\footnote{e-mail: \tt ftlipkin@weizmann.ac.il} \\
\vskip0.8cm
{\it
$^b\;$School of Physics and Astronomy \\
Raymond and Beverly Sackler Faculty of Exact Sciences \\
Tel Aviv University, Tel Aviv, Israel\\
\vbox{\vskip 0.0truecm}
$^c\;$Department of Particle Physics \\
Weizmann Institute of Science, Rehovot 76100, Israel \\
and\\
High Energy Physics Division, Argonne National Laboratory \\
Argonne, IL 60439-4815, USA
}
\end{center}

\vspace*{0.8cm}

\centerline{\bf Abstract}

\vspace*{4mm}

Observation of CP violation in
$B^o\rightarrow K^\pm\pi^{\mp}$ decays and its absence in $B^+\rightarrow K^+\pi^o$
decays are explained in new improved data analysis of more precise $B\rightarrow K\pi$ data.
Success of the ``Lipkin Sum Rule" indicates that four $B\rightarrow K\pi$ branching ratios
are determined by three parameters, the penguin diagram $P$ and two interference terms $P\cdot T$
and $P\cdot S$ between the dominant penguin and two tree diagrams; the color-favored and color suppressed
diagrams. Previous analyzes confirmed the model with errors leaving values
of interference terms less that two standard deviations from zero. The observation CP violation in
$B^o\rightarrow K^\pm\pi^{\mp}$ decays indicates a finite value for $P\cdot T$. New precise data analysis show
$P\cdot T$ and $P\cdot S$ interference contributions well above errors.
Their contributions
to $B^\pm\rightarrow K\pi$ decays are shown to be nearly equal with opposite phase and cancel within experimental errors. This cancelation unexpected in previous analyzes explains the failure to
see CP violation in $B^\pm\rightarrow K\pi$ decays. It can be due to the Pauli antisymmetry exchange neglected in previous analyzes. Two $B^\pm\rightarrow K\pi$ tree diagrams differ by interchange of two identical $u$ quarks. $B^o\rightarrow K^\pm\pi^{\mp}$ diagrams have no identical quark pairs. This Pauli effect explains the
difference  produced by changing the flavor of the spectator quark which does not participate in the weak interaction. Our analysis differs from previous analyzes which assume SU(3) flavor symmetry to use input from $B\rightarrow \pi\pi$ data and neglect Pauli effects.
We use new data, include  Pauli effects and strong final state interactions to all orders in QCD
with no higher flavor symmetry assumed beyond isospin.
We do not use $B\rightarrow \pi\pi$ data.

\vfill\eject

\section {Introduction}

A general theorem from CPT invariance shows\cite{lipCPT} that direct CP
violation can occur only via the interference between two amplitudes which have
different weak phases and different strong phases. This  holds also for all
contributions from new physics beyond the standard model which conserve CPT.

Direct CP violation has been experimentally observed\cite{PDG,HFAG}
in $B_d \rightarrow K^+ \pi^-$ decays.
\beq{acp0}
A_{CP}(B_d \rightarrow K^+ \pi^-)= -0.098 \pm 0.013
\eeq
This CP violation has been attributed to the interference between the large contribution from the dominant penguin diagram and smaller contributions from tree diagrams.  The failure to observe CP violation in charged
decays\cite{Ali} is has been considered to be a puzzle\cite{nurosgro,ROSGRO}
because changing the flavor of a spectator quark which does not participate
in the weak decay vertex is not expected to  make a difference.
 \beq{acp+}
\begin{array}{ccl}
\displaystyle
A_{CP}(B^+ \rightarrow K^o_S \pi^+)= 0.009 \pm 0.029
\hfill\\
\\
A_{CP}(B^+ \rightarrow K^+ \pi^o)= 0.051 \pm 0.025
\end{array}
\end{equation}
We shall show here that the dependence on spectator flavor arises from the Pauli blocking
by the spectator quark of a quark of the same flavor participating in the weak vertex. The quark produced by a
tree diagram is a  u-quark  which is Pauli blocked by the spectator $u$ quark in $B^+$ decay but 
is not affected by the spectator $d$ quark in neutral decays. This difference in Pauli 
blocking suppresses the tree contribution and CP violation in charged $B$ decays but allows
tree-penguin interference and enables CP violation to be observed in neutral decays. 
\section{The crucial role of Pauli blocking by the spectator quark}
\subsection{The difference between charged and neutral decays}
The Pauli principle can forbid the tree-penguin interference and CP
violation in  charged B decays and allow them in neutral  decays.
The decay of a $\bar b$ antiquark to a strange charmless final state is described by the
vertex having the form
\beq{bvert}
\bar b \rightarrow
\bar s n \bar n
\end{equation}
where $n \bar n$ denotes a nonstrange $u \bar u$ or $d \bar d$. Pauli blocking
suppresses the transition for a state $n$ which has the same flavor as the spectator quark.
This suppression is lost in conventional treatments which consider color-favored and color-suppressed
tree amplitudes as independent without considering Pauli suppression. We show below that it is just this
Pauli selection rule forbidding a $u$ quark produced by a weak interaction to enter the same state as a
$u$ spectator quark which suppresses the tree contribution and CP
violation in  charged B decay.

As a simple approximation we neglect the color and spin degrees of freedom. Corrections from color-spin effects will be considered later.
Let $u^\dag$, $d^\dag$ and $Q^\dag$ denote creation operators respectively for a $u$ quark, a $d$ quark and a spectator quark denoted by $Q$ and let $\bar b^\dag$, $\bar u^\dag$,  $\bar d^\dag$ and $\bar Q^\dag$ denote creation operators respectively for a $\bar b$ antiquark,  $\bar u$ antiquark, a $\bar d$ antiquark and a spectator $\bar Q$ antiquark.

The transition from an initial $B$ meson state $\bar b^\dag\ Q^\dag\ket{0}$ consisting of a $\bar b$ antiquark and nonstrange quark to a strange charmless two-meson final state is written
\beq{secquan}
\ket{B}=\bar b^\dag\ Q^\dag\ket{0} \rightarrow
\bar s^\dag  \cdot \left[d^\dag \bar d^\dag + u^\dag \bar u^\dag +\xi \cdot u^\dag \bar u^\dag\right] Q^\dag\ket{0}
\end{equation}
where the final state is written as the sum of an isoscalar $q\bar q$ pair and a $u\bar u$ pair together with a strange antiquark and a spectator quark. This is analogous to conventional description as the sum of a penguin contribution and a tree contribution with a parameter $\xi$ generally considered to be small expressing the ratio of the tree and penguin contributions. Substituting the two spectator flavors $u$ and $d$ then gives
\beq{secquan2}
\begin{array}{ccl}
\displaystyle
{\bar b^\dag\ u^\dag}\ket{0} \rightarrow
\bar s^\dag  \cdot \left[d^\dag \bar d^\dag + (1+\xi) u^\dag \bar u^\dag\right] u^\dag\ket{0}=
\bar s^\dag  \cdot d^\dag \bar d^\dag u^\dag\ket{0}=\left[K^o \pi^+ + \frac {K^+ \pi^o}{\sqrt 2} \right]
\hfill\\
{\bar b^\dag\ d^\dag}\ket{0} \rightarrow
\bar s^\dag  \cdot \left[d^\dag \bar d^\dag + (1+\xi) u^\dag \bar u^\dag\right] d^\dag\ket{0}=(1+\xi)
\bar s^\dag  \cdot u^\dag \bar u^\dag d^\dag\ket{0}=\left[K^+ \pi^- + \frac {K^o  \pi^o}{\sqrt 2} \right]
\cdot (1+\xi)
\end{array}
\end{equation}
where we have noted that the products of two identical fermion creation operators $d^\dag d^\dag$ and $u^\dag u^\dag $
must vanish. The transitions for the neutral decays are seen to depend upon the parameter $\xi$ while the charged transitions are seen to  be independent of the parameter $\xi$. The parameter $\xi$ is proportional to the strength of the tree amplitude. Thus tree-penguin interference which might produce CP violation is present in neutral decays and absent in charged decays. This explaina how CP violation can be drastically changed by changing the spectator quark
and the otherwise mysterious result  (\ref{acp+}).
\subsection{The Lipkin Sum Rule}
The transition matrices denoted by $T$ must satisfy the isospin constraints for the coupling of an isospin 1 meson to an isospin (1/2) kaon
\beq{secquanamp2}
\begin{array}{ccl}
\bra{K^o \pi^o}T \ket{B^o} =- \frac {1}{\sqrt 2}\cdot \bra{K^+ \pi^-}T \ket{B^o}; ~ ~ ~ ~
2\cdot |\bra{K^o \pi^o}T \ket{B^o}|^2=
|\bra{K^+ \pi^-}T \ket{B^o}|^2
\hfill\\
\bra{K^+ \pi^o}T \ket{B^+} =- \frac {1}{\sqrt 2}\cdot \bra{K^o \pi^+}T \ket{B^+}; ~ ~ ~ ~
2\cdot |\bra{K^+ \pi^o}T \ket{B^+}|^2=
|\bra{K^o \pi^+}T \ket{B^+}|^2
\end{array}
\end{equation}
Then
\beq{secquanamp3}
\frac {\bra{K^+ \pi^-}T \ket{B^o}}{\bra{K^o \pi^+}T \ket{B^+}}=(1+\xi)=
\frac {\bra{K^o \pi^o}T \ket{B^o}}{\bra{K^+ \pi^o}T \ket{B^+}}
\end{equation}
\beq{eqapp3}
2\cdot |\bra{K^+ \pi^o}T \ket{B^+}|^2-
|\bra{K^o \pi^+}T \ket{B^+}|^2
=0=
(1+\xi)^2\cdot\left[|\bra{K^+ \pi^-}T \ket{B^o}|^2-2|\bra{K^o \pi^o}T \ket{B^o}|^2\right]
\eeq
this can be rewritten

\beq{eqapp4}
2\cdot \left[|\bra{K^+ \pi^o}T \ket{B^+}|^2+|\bra{K^o \pi^o}T \ket{B^o}|^2 \right]
=|\bra{K^o \pi^+}T \ket{B^+}|^2+
|\bra{K^+ \pi^-}T \ket{B^o}|^2
\eeq
This relation (\ref{eqapp4}) is  seen to be identical to the approximate equality\cite{approxlip,approxgr} called the ``Lipkin sum rule"\cite{Ali}.

\beq{sumruleapp}
R_L \equiv 2{{\Gamma(B^+ \rightarrow K^+ \pi^o) + \Gamma(B^o
\rightarrow K^o \pi^o)} \over {\Gamma(B^+ \rightarrow K^o \pi^+ )
+  \Gamma(B^o
\rightarrow K^+ \pi^-)}} \approx 1
\eeq

\section {Three parameters  determine  four $B\rightarrow K\pi$ branching ratios }
\subsection {Two independent derivations of the sum rule}
The sum rule (\ref{sumruleapp}) has been derived here using a completely different set
of assumptions from the previous derivation \cite{approxlip,approxgr,Gronau,ketaprimfix}.
Both derivations begin with the relation (\ref{secquan})
The derivation of the relation (\ref{eqapp4}) neglects the color and spin degrees of freedom but makes
no further assumptions and includes Pauli blocking.
The experimental observation (\ref{acp0}) and the knowledge that the penguin
amplitude is dominant for the decay\cite{Ali} require that the decay amplitude must
contain at least one additional amplitude with both weak and strong phases
different from those of the penguin.

The standard treatment assumes that four $B\rightarrow K\pi$ branching ratios
are determined by three parameters, the penguin diagram $P$ shown in Fig. 1 and two interference terms $P\cdot T$
and $P\cdot S$ between the dominant penguin and the color-favored and color suppressed tree
diagrams shown in Figs. 2 and 3. This derivation assumes the difference between the two three contributions shown in Figs. 2 and 3 are independent and neglects Pauli blocking. It also assumes that the two tree amplitudes are sufficiently small to be
treated in first order. Second order terms $T\cdot T$, $S\cdot T$
and $S\cdot S$ are shown to be negligible.
The agreement\cite{Ali} with experiment\cite{PDG,HFAG} confirms either these assumptions  \cite{approxlip,approxgr,Gronau,ketaprimfix} or the neglect of spin and color while including Pauli blocking.
\subsection {The Difference rule}
We now investigate what is observable in the experimental data, how to separate
the signal from the noise, how to find the`other amplitude and examine what can we learn
about it from experiment.
The sum rule (\ref{sumruleapp}) has been rearranged \cite{ketaprimfix}
to obtain a ``difference rule"
\beq{eqapp}
2B(B^+ \rightarrow K^+ \pi^o)
- B(B^+ \rightarrow K^o \pi^+ ) \approx
B(B^o \rightarrow K^+ \pi^-)  - 2B(B^o\rightarrow K^o \pi^o)
\eeq
where for simplicity the result was expressed in terms of  branching ratios, denote
by B(). Corrections for the difference between the $B^+$ and $B^o$ lifetime ratio are
included in the subsequent analysis

The difference rule (\ref{eqapp}) has real
significance only if there is interference between the dominant penguin and
another amplitude leading to an I=3/2 final state. It is trivially satisfied if
the decays are described entirely by a pure penguin or other I=1/2 contribution
where the final state is an isospin eigenstate with I=1/2. For a pure penguin
transition both sides of the difference rule (\ref{eqapp}) vanish and the
relation reduces to the trivial 0=0.

Four experimental branching ratios
for  $B \rightarrow K\pi$ are available\cite{PDG,HFAG}. Three different independent
differences between these branching ratios can be defined which eliminate the
penguin contribution. These overdetermine the two remaining free parameters
in the theory, the interference contributions between the penguin amplitude and
the two tree amplitudes.
The next step in the analysis would have been to solve the equations and obtain the
values of the two interference terms $P\cdot T$
and $P\cdot S$  from the experimental data. However the large experimental errors at the time
left these values less that two standard deviations from zero\cite{approxlip,approxgr}.

\subsection{New experimental data show statistically significant penguin-tree interference}

New data now show  that the $B\rightarrow K\pi$ transition is not a pure penguin. They give
a finite experimental value for an expression which vanishes in a pure penguin transition.
\beq{newpuz2}
{{\tau^o}\over{\tau^+}}\cdot 2B(B^+ \rightarrow K^+ \pi^o) -
B(B^o \rightarrow K^+ \pi^- ) =
4.7 \pm 0.82
\eeq
where $\tau^o/\tau^+$ denotes the ratio of the $B^o$ and $B^+$ lifetimes and we have used
the experimental values
\beq{newtestex}
\begin{array}{ccl}
\displaystyle
B(B^o \rightarrow K^+ \pi^-)  =
19.4 \pm 0.6
 \hfill\\
\\
\displaystyle
{{\tau^o}\over{\tau^+}}\cdot B(B^+ \rightarrow K^o \pi^+) =
%
{{(23.1 \pm 1.0)}\over{1.07}}=
21.6\pm 0.93
\hfill\\
\\
\displaystyle
{{\tau^o}\over{\tau^+}}\cdot 2B(B^+ \rightarrow K^+ \pi^o) =
{{2\cdot (12.9\pm 0.6)}\over{1.07}}=
24.1\pm 0.56
\hfill\\
\\
\displaystyle
B(B^o\rightarrow K^o \pi^o) =
(9.4 \pm 0.6)
\end{array}
\end{equation}

\subsection{A surprising cancelation suggests Pauli effects $P\cdot (T + S) \approx 0 $}

We now note that the right hand side of eq.(\ref{eqapp}) gives
\beq{newpuz}
 B(B^o \rightarrow K^+ \pi^-) - 2B(B^o\rightarrow K^o \pi^o) =
 (19.4 \pm 0.6)- 2\cdot (9.4 \pm 0.6) = 0.6 \pm 1.3
\eeq

The expression (\ref{newpuz}) also vanishes in the case of a pure penguin transition.
The significant difference between the experimental values of expressions (\ref{newpuz}) and (\ref{newpuz2})
which both vanish in the case of a pure penguin transition seems to indicate a surprising cancelation and
motivates a search for a theoretical explanation.

We first note that the two transitions (\ref{newpuz})  which have a d-quark spectator are seen to be described
respectively by the color-favored and color-suppressed tree diagrams shown respectively in
figs. 2 and 3 in addition to the dominant common penguin diagram  shown in
fig. 1 . This
cancelation between the contributions of the two tree diagrams is surprising because
the standard treatments assume that the these two tree contributions
are completely independent and are not expected to cancel.
A further analysis of the new data \cite{PDG,HFAG}
isolates the color-favored and color-suppressed  contributions,
and minimizes the importance of experimental errors.
\beq{newtest0}
\frac{\vec P\cdot (\vec T + \vec S)}{\vec P\cdot (\vec T - \vec S)}
=\frac{2B(B^o\rightarrow K^o \pi^o) - B(B^o \rightarrow K^+ \pi^-)} {{{[\tau^o}/{\tau^+]}}\cdot[ B(B^+ \rightarrow K^o \pi^+) + 2B(B^+ \rightarrow K^+ \pi^o)] -
2B(B^o \rightarrow K^+ \pi^- )}
=0.09\pm 0.1
\end{equation}

We now note that the Pauli principle  neglected  in conventional treatments
can produce this cancelation. The amplitudes $T$ and $S$  go into one another under
the interchange of the two identical $u$ quarks in $A[K^+\pi^o]$.
A full examination
of Pauli effects requires antisymmetrization of the $uu$ wave function including the
color and spin correlations. As a  first approximation we neglect color and spin.
Then Pauli antisymmetry
requires $T$ and $S$ amplitudes to be equal and opposite and explains the cancelation
(\ref{newtest0}).
Both $B^+$ decays are then pure penguin decays to the $I=1/2$ $K\pi$ state.
Experiment\cite{HFAG} shows  agreement with this prediction to  between one
and two standard deviations.

\beq{bplusselec2}
2B(B^+ \rightarrow K^+ \pi^o) = 25.8\pm 1.2 \approx
B(B^+ \rightarrow K^o \pi^+ ) = 23.1 \pm 1.0
\eeq
where $B$ denotes the branching ratio in units of $10^-6$

The data are now sufficiently precise to show  that
the interference terms between the dominant penguin amplitude
and the two tree amplitudes are both individually
finite and one is well above the experimental errors. The sum
rule is satisfied and is now nontrivial. But the interference term
$\vec P \cdot (\vec T+\vec S)$ is now equal to zero
well within the experimental errors (\ref{newtest0}). This confirms
the Pauli symmetry prediction
(\ref{bplusselec2}) with smaller experimental errors.

Thus tree-penguin interference with normally ignored Pauli effects
can explain the observed CP violation in
charged B-decays and its absence in neutral decays.

This shows how a nontrivial change in the weak
decay amplitude can arise from a change of the flavor of the spectator quark.

\subsection {A new analysis of the data pinpointing tree-penguin interference}

We now show how a detailed conventional analysis of new experimental data with no new theory
leads to the result (\ref{newtest0}). We later show how this cancelation can
result from the Pauli principle.

The conventional analysis expresses the four  $B \rightarrow K\pi$
amplitudes in terms of the three amplitudes $P$, $T$ and $S$
denoting respectively the penguin,  color favored tree and color suppressed
tree amplitudes while neglecting other contributions at this
stage\cite{approxlip,approxgr,Gronau,ketaprimfix}

\beq{pst}
\begin{array}{ccl}
\displaystyle
A[K^o\pi^+]=P; ~ ~ ~ A[K^+\pi^-]= T + P
\hfill\\
\\
\displaystyle
A[K^o\pi^o]={{1}\over{\sqrt{2}}} [S -  P]; ~ ~ ~
A[K^+\pi^o]={{1}\over{\sqrt{2}}} [T + S + P]
\end{array}
\end{equation}

When the interference terms are taken only to first order,

\beq{pstsq}
\begin{array}{ccl}
\displaystyle
|A[K^o\pi^+]|^2
=|\vec {{ P}}|^2; ~ ~ ~ |A[K^+\pi^-]|^2 =|\vec {{ P}}|^2+2\vec { P} \cdot \vec T
\hfill\\
\\
\displaystyle
2\cdot |A[K^o\pi^o]|^2
=|\vec {{ P}}|^2  -\vec  P \cdot \vec S; ~ ~ ~ 2\cdot |A[K^+\pi^o]|^2
=|\vec {{ P}}|^2+2\vec { P} \cdot (\vec T+\vec S)
\end{array}
\end{equation}
where the approximate equalities hold to first order in the $T$ and $S$
amplitudes.

We now get a more sensitive experimental test by using all the $B \rightarrow K\pi$ data.

We  use new
data and define new differences which optimize the signal to noise ratio.
Noting that the branching ratio $B(B^o \rightarrow K^+ \pi^-) $ has the
smallest experimental error, we define three independent differences which
vanish for a pure penguin transition and are chosen to have the smallest
experimental errors.

\begin{equation}
\begin{array}{ccl}
\displaystyle
\Delta (K^o\pi^+) \equiv |A[K^o\pi^+]|^2 - |A[K^+\pi^-]|^2 \approx - 2
\vec {{ P}} \cdot \vec T
\hfill\\
\\
\displaystyle
\Delta (K^+\pi^o) \equiv 2 |A[K^+\pi^o]|^2 - |A[K^+\pi^-]|^2 \approx 2
\vec { P} \cdot \vec S
\hfill\\
\\
\displaystyle
\Delta (K^o\pi^o) \equiv 2 |A[K^o\pi^o]|^2 - |A[K^+\pi^-]|^2 \approx - 2
\vec  P \cdot (\vec T+\vec S)
\end{array}
\end{equation}
where the appoximate equalities hold to first order in the $T$ and $S$
amplitudes.
The isospin sum
rule\cite{approxlip,approxgr} is easily expressed in terms of these differences,

\beq{eqapprev}
\Delta (K^o\pi^o) + \Delta (K^+\pi^o) - \Delta (K^o\pi^+)
  \approx 0
\eeq

Since each of the three terms in eq. (\ref{eqapprev}) vanish for a pure penguin
transition, the sum rule is trivially satisfied in this case. We
improve the previous analysis\cite{ketaprimfix}  that only showed
the sum rule trivially satisfied with real data and
all terms proportional to tree-penguin interference were still statistically
consistent with zero.

These individual differences are now sufficiently different
from zero with available experimental branching ratio data corrected for the
lifetime ratio\cite{PDG,HFAG}

\beq{newtest}
\begin{array}{ccl}
\displaystyle
{{\tau^o}\over{\tau^+}}\cdot B(B^+ \rightarrow K^o \pi^+) -
B(B^o \rightarrow K^+ \pi^-)  =
2.2 \pm 1.1 \propto -\vec  P \cdot \vec T
\hfill\\
\\
\displaystyle
{{\tau^o}\over{\tau^+}}\cdot 2B(B^+ \rightarrow K^+ \pi^o) -
B(B^o \rightarrow K^+ \pi^- ) =
4.7 \pm 0.82
\propto\vec P \cdot \vec S
\hfill\\
\\
\displaystyle
2B(B^o\rightarrow K^o \pi^o) - B(B^o \rightarrow K^+ \pi^-) =
-0.6 \pm 1.3
\propto - \vec P \cdot (\vec T + \vec S)
\end{array}
\end{equation}

Combining these equations gives the relation (\ref{newtest0}).

There is no new theory here.
Choosing three independent differences in a way to minimize experimental errors
shows two significant signals  well
above the noise of experimental errors that still fit an
overdetermination of the two parameters and lead to the result (\ref{newtest0}). These show two finite
tree-penguin interference contributions that can produce the observed
direct CP violation in neutral B-decays.  However the third difference is
consistent with zero well below the noise and below the other two
contributions. The absence of tree-penguin contributions in this difference is
completely unpredicted in the standard treatments.

\section{Symmetry arguments supporting the vanishing of $\vec  P \cdot (\vec T + \vec S)$}
\subsection {The $u n \bar u \bar s$ tree diagrams for $B \rightarrow K \pi$ decays}

In tree diagrams for charmless strange $B$ decays, the $\bar b \rightarrow u \bar u \bar s$ transition produces a four-quark state $u n \bar u \bar s$ where $n$ denotes the nonstrange spectator quark.

The $K^+\pi^-$ final state $(u  \bar s)(n \bar u)$ can be produced by a $B^o$ tree diagram in which the spectator quark $n$ is a $d$ quark and combines with the $\bar u$ in a color-favored transition shown in Fig. 2.
The CP violation observed in this state indicates that is produced by appreciable $P\cdot T$ interference.

The $K^+\pi^o$ final state $(u  \bar s)(n \bar u)$ or $(n  \bar s)(u  \bar u)$ can be produced by a $B^+$ tree diagram in which the spectator quark $n$ is a $u$ quark  and combines with either the $\bar u$ in a color-favored transition
shown in Fig. 2
or the $\bar s$ in a color-suppressed transition shown in Fig.3. The failure to observe CP violation in this state while CP violation is observed in the $K^+\pi^-$ final state indicates that both $P\cdot T$
and $P\cdot S$ interference contributions are appreciable and their interference contributions have opposite phase and tend to cancel any CP violation.

The experimental data for these two transitions thus present predictions for the following two final states.

The $K^o\pi^o$ final state $(n  \bar s)(n \bar u)$  can be produced by a $B^o$ tree diagram  in which the spectator quark $n$ is a $d$ quark and combines with the $\bar s$ in a color-suppressed transition shown in Fig. 3. This transition is produced by $P\cdot S$ interference  which is expected to be similar to the  $P\cdot T$
interference contribution and produce a similar CP violation to that  observed in
the $K^+\pi^-$ final state

The $K^o\pi^+$ final state $(d\bar s)(u\bar d)$ contains a $\bar d$ antiquark and cannot be produced by a tree diagram leading to a $u n \bar u \bar s$ state.
The prediction for the transition to
this final state is that it has no tree contribution, no penguin-tree interference and no CP violation.

\subsection{Experimental results suggest underlying symmetry}

Experimental results for $B^o$ decays now show finite penguin-tree interference and a possibility of  CP violation.
Results for $B^{\pm}$ decays show negligible penguin-tree interference and little possibility of CP violation.

We now show how these results follow from symmetry conditions.

In charged $B$ decays the spectator quark is a $u$ quark and the $un$ pair has s unique isospin $I=1$. The two identical fermions must obey Pauli antisymmetry.
In neutral $B$ decays the spectator quark is a $d$ quark, the $un$ pair is a mixture of two isospins $I=1$ and $I=0$ and has no Pauli constraints.

A full analysis of symmetry and Pauli effects must include color-spin correlations.
The state of two u quarks in a relative S wave which are
symmetric in space and flavor must be antisymmetric in color and spin.
A group-theoretical treatment of this symmetry involves the color-spin
SU(6) group in which all pseudoscalar mesons are color-spin singlets, while the uu
diquark in a relative s-wave is classified in the antisymmetric 15 dimensional
representation.

\subsection{Isospin Analysis}

We first examine Pauli effects in the present data using only isospin
and see how these produce a selection rule that cancels the
tree contribution to $B^+ \rightarrow K^+\pi^o$.

The tree diagram for $B^+ \rightarrow K^+\pi^o$ has
a four-body  $u\bar s u\bar u$ state containing a $u$ spectator quark and the
$\bar u u \bar s $ produced
by the $\bar b$ antiquark weak decay.  This state contains two $u$ quarks with
isospin $I=1,I_z=+1$. The $\bar u$ antiquark is in a well  defined isospin state
with $I=1/2,I_z=-(1/2)$ and the strange antiquark has  isospin zero. The total
four-body state is thus a state with well defined isospin.It is a definite mixture
of two eigenstates of the total four-body isospin with $I=1/2$ and $I=3/2$ with
unique relative magnitudes and phases with determined by
isospin Clebsch-Gordan coefficients for coupling two
states with $I=1$ and $I=1/2$ to  $I=1/2$ and $I=3/2$..

\beq{initial}
\ket{i;u\bar s u\bar u} \propto \ket{{1\over 2},{1\over 2}}
\bra{1{1\over 2}1(-{1\over 2})}
{1{1\over 2}{1\over 2}{1\over 2}}  +
{{3\over 2},{1\over 2}}
\rangle
\bra{1{1\over 2}1(-{1\over 2})}
{1{1\over 2}{3\over 2}{1\over 2}}
\rangle
\eeq
where $\langle {j_1j_2m_1m_2}\ket{j_1j_2JM}$ denotes a Clebsch-Gordan coefficient.

The $K\pi$ final states are also linear combinations of states with isospin (1/2) and
isospin (3/2) with relative amplitudes and phases determined completely by the
requirement that the pion has isospin one and the kaon has isospin (1/2).

\beq{finalkp}
\ket{f;\pi^oK^+} \propto \ket{{1\over 2},{1\over 2}}
\bra{ {1{1\over 2}0({1\over 2})}}
{1{1\over 2}{1\over 2}{1\over 2}}  +
{{3\over 2},{1\over 2}}
\rangle
\bra{1{1\over 2}0({1\over 2})}
{1{1\over 2}{3\over 2}{1\over 2}}
\rangle
\eeq

 From the orthogonality relation for Clebsch-Gordan coefficients
\beq{ortho}
\langle f;\pi^oK^+ \ket{i;u\bar s u\bar u} =0.
\eeq

There is therefore no overlap between the initial state
$\ket{i;u\bar s u\bar u}$ produced by the weak interaction tree diagram
and the $K^+\pi^o$ final state. The relevant Clebsch-Gordan coefficients
for the $K^+\pi^o$ state  are seen to be
just those to make this linear combination exactly orthogonal to the
combination in the $u\bar s u\bar u$ state produced in the tree diagram
by the $\bar b$ decay .
The overlap between the $K^+\pi^o$ state and the initial state thus vanishes.
In the ``fall-apart"\cite{barnes} model commonly used in tetraquark decays
this vanishing overlap indicates that  the the tree diagram contribution to
$B^+ \rightarrow K^+\pi^o$ transition vanishes.

The transition is therefore forbidden if the four-body state that fragments
into a kaon and a pion $\ket{u\bar s u\bar u}$ retains the $u\bar s u\bar u$ constituents with
the uu pair coupled to I=1. This is the same as the original four-body state created in
the tree diagram by the $\bar b$ antiquark decay. This is true in the simple
``fall-apart"\cite{barnes} model. The
experimental data seem to indicate that the tree diagram for this transition
is indeed forbidden here.

The only way that the initial $\ket{u\bar s u\bar u}$ state can change by a
strong interaction conserving isospin is annihilating the $u\bar u$ pair and creating a $d\bar d$
pair. This adds a $\ket{u\bar s d\bar d}$ component to the final state.
This interaction is included below in a general treatment including color-spin correlations

We now note that changing the flavor of the spectator quark makes a crucial
difference in the isospin analysis. The isospin of the two quarks $(u,d)$ in the corresponding tree diagram for
the neutral $B$ decays is not unique, it is a combination of $I=0$ and $I=1$.
Thus there is no  isospin nor Pauli constraint here and no selection rule forbidding the
tree contribution.

\subsection {Detailed symmetry and Pauli analysis}

We now examine the effect of symmetry restrictions from the Pauli principle on
the fragmentation of a $uu\bar u \bar s$ state into a $K^+\pi^o$ color singlet
state with spin zero and no orbital angular momentum.  Explicitly writing wave
functions and imposing
Pauli antisymmetry requires the full color-spin-flavor SU(12)
group. Since the $uu$ state
is flavor symmetric it must be
antisymmetric in color and spin if it is in a spatially symmetric S-wave. The
fragmentation process is a strong interaction which conserves flavor SU(3) and
charge conjugation. Since the initial state contains no $d$ quarks we can
simplify the symmetry restriction by considering the $V$-spin subgroup of SU(3)
which acts in the $u-s$ flavor space.
This state contains a  $uu$ pair required by the Pauli principle to be
antisymmetric in color and spin. It is either in a color antitriplet with spin
$S=1$ or in a color sextet with spin $S=0$.  The  $\bar u \bar s$ pair must
also be either in a color antitriplet with spin $S=1$ or in a color sextet with
spin $S=0$. Although no Pauli principle forbids it from being in a color
antitriplet with spin $S=0$ or in a color sextet with spin $S=1$ these states
cannot combine with the $uu$ pair to make the spin-zero color singlet final
state.

Both the $uu$ diquark and the $\bar u \bar s$ antidiquark are thus
antisymmetric in color and spin. The generalized Pauli principle requires each
to be symmetric in flavor SU(3) and its  SU(2) subgroup V-spin. Each is therefore
in the symmetric V-spin state with $V=1$.
A final state must be even under  general charge conjugation to decay into two
pseudoscalar mesons in an orbital S wave. Thus the $(V=1,V_z=+1)$ diquark and
the $(V=1,V_z=0)$ antidiquark must be coupled symmetrically to  $(V=2,V_z=+1)$.
This state is in the 27-dimensional representation of flavor SU(3).
The $V$ spin analysis of the initial state $\ket{i;uu\bar d \bar s}$ is
\beq{initialuuds}
\begin{array}{ccl}
\displaystyle
\ket{i;uu\bar d \bar s} = \frac {1}{\sqrt 2}\cdot  \ket {V=1;V_z=1} \cdot
\left [\ket {V=1;V_z=0} + \ket {V=0;V_z=0}\right]
\hfill\\
\\
\bra{i;uu\bar d \bar s}{V=2;V_z=1}\rangle =\frac{1}{2}
\end{array}
\end{equation}

A final $K^o\pi^+$ state $\ket{f;\pi^oK^+}$
has no $V=2$ component, since both the $K^o$ and $\pi^+$
have V=1/2. Thus the tree diagram for the $K^o\pi^+$ decay must vanish and this
decay is pure penguin, in agreement with the isospin analysis (\ref{pst})

The final $K^+\pi^o$ state contains a $\pi^o$ which is a linear combination of a
$u\bar u$ pair and a $d\bar d$ pair. Since the $d$ quark has $V=0$ the $d\bar d$ pair
has $V=0$  and cannot combine with a $V=1$ $K^+$ to make $V=2$. The $u\bar u$ pair is an
equal mixture of $V=0$  and $V=1$. Only the $V=1$ component can   combine with a $V=1$ to make $V=2$.
Thus the V-spin analysis of the final
$K^o\pi^+$ state $\ket{f;\pi^oK^+}$ and its overlap with the initial state
are
\beq{finalkp}
\begin{array}{ccl}
\displaystyle
\bra{{K^+\pi^o}}{V=2;V_z=1}\rangle  =\frac{1}{4}
\hfill\\
\\
\bra{{K^+\pi^o}}i;uu\bar d\rangle  =\frac{1}{8}
\end{array}
\end{equation}

We thus see that the tree diagram for transition $B^+\rightarrow K^+\pi^o$ state
vanishes in the ``fall-apart" model where the initial $u\bar u$ pair does not
have enough time to annihilate and create a $d\bar d$ pair. When there is time
for the $u\bar u$ pair annihilation and the creation a $d\bar d$ pair, the
tree diagram does not vanish but is suppressed by a significant factor. Present
data are consistent with complete suppression but evidence for a partial suppression
is still down in the noise.

The $ud\bar u \bar s$ state created in the tree diagram for
$B_d$ decay has no such restrictions. It can be in a flavor SU(3)
octet as well as a 27. Its ``diquark-antidiquark" configuration includes the
flavor-SU(3) octet constructed from the spin-zero color-antitriplet
flavor-antitriplet ``good" diquark found in the  $\Lambda$ baryon and its
conjugate ``good" antidiquark. These ``good diquarks" do not exist in the
corresponding $uu\bar u \bar s$ configuration.

We again see that the Pauli effects produce a drastic dependence on spectator
quark flavor in the tree diagrams for $B \rightarrow K \pi$ decays.

Thus tree-penguin interference can explain both the presence
of CP violation in neutral decays and its absence charged decays.


\section{A flavor topology analysis which includes final state interactions}
The unique flavor topology of the charmless strange quasi-two-body weak B decays
enables the results (\ref{newtest}) to be obtained in a more general analysis
of these decays
including almost all possible diagrams including final state interactions and complicated
multiparticle intermediate states.

Consider diagrams for a charmless $B(\bar b q_s)$ decay into one strange and
one nonstrange meson, where $q_s$ denotes either a $u$ or $d$. The allowed
final  states must have the quark constituents  $\bar s n \bar n q_s$ where $n$
denotes a $u$ or $d$ nonstrange quark. We consider the topologies of all
possible diagrams in which a $\bar b$ antiquark and a nonstrange quark enter a
black box from which two final $q \bar q$ pairs emerge. We follow the quark
lines of the four  final state particles through the diagram going backward and
forward in time until they reach either the initial state or  a vertex where
they are created. There are only two possible quark-line topologies for these
diagrams:

\begin{enumerate}

\item We call a generalized penguin diagram, shown in Fig. 1 , the sum of all possible diagrams in which a $\bar q q$ pair appearing in
the final state  is created by a gluon somewhere in the diagram. The quark
lines for the  remaining pair must go back to the weak vertex or the initial
state. This diagram includes not only the normally called penguin diagram but
all other diagrams in which the final  pair is created by gluons somewhere in
the diagram. This includes for example all diagrams normally called ``tree
diagrams" in which an outgoing $u \bar u$ or $c \bar c$ pair  is annihilated into gluons in a
final state interaction and a new isoscalar  $\bar q q$ pair is created by the
gluons.
There are two topologies for penguin diagrams.
\begin{itemize}
\item. A normal penguin diagram has a the spectator quark line continuing unbroken
 from the initial state to the final state.
 This  penguin contribution is described by a single parameter,
denoted by $P$ which is independent of the spectator quark flavor and contributes
equally to the $\bar s u \bar u q_s$ and $\bar s d \bar d q_s$ states.
\item A diagram which we call here an ``anomalous penguin" has the spectator
``u" quark in a $B^+$ decay annihilated in a final state interaction
against the $\bar u$ antiquark produced in a tree diagram. This diagram also   contributes
equally to the $\bar s u \bar u q_s$ and $\bar s d \bar d q_s$ states. But this
diagram denoted by $P_u$ is present only in
charged decays.
\end{itemize}
\item We call the ``tree diagram" the sum of all possible diagrams in which all
of the four quark lines leading to the final state go back to a initial $\bar s
u \bar u$ state created by the weak decay of the $b$ quark and the $ q_s$
spectator whose line goes back to the initial state.
     There are two possible couplings of the pairs to create final two meson
states from this diagram
 \begin{itemize}

\item The $\bar s u$ pair is coupled to make a strange meson; the  $\bar u q_s$
pair is coupled to make a nonstrange meson as shown in Fig.2. This is
conventionally called the  color-favored coupling. The contribution
of this coupling is described by a  single parameter, denoted by $T$.

\item The $\bar s q_s$ pair is coupled to make a strange meson; the  $u \bar u$
is coupled to make a nonstrange meson as shown in Fig. 3.
This is conventionally called the  color-suppressed coupling. The contribution
of this coupling is described by a  single parameter, denoted by $S$.

 \end{itemize}
 \end{enumerate}

All the results (\ref{newtest}) obtained with the conventional definitions of $P$, $T$ and $S$ are seen to hold here
with the new definitions of $P$, $T$ and $S$. They now include contributions from
all final state interactions which conserve isospin and do not change
quark flavor.
The one final
state interaction not included is the $P_u$ diagram occurring in $B^+$ decays.
The flavor topology of this diagram creates an additional $I=1/2$ state
which is neglected in the derivation of the results (\ref{newtest}). These results hold as
long as the contribution of this $P_u$ diagram by final state interactions to the observed final states is negligible.
That this additional $I=1/2$ contribution does not affect the sum rule (\ref{sumruleapp})
is easily seen by its representation
as a ``difference rule" (\ref{eqapp}) which considers only the $I=3/2$ contributions.

In neutral $B_d$ decays there is no $P_u$ diagram.
Thus the simple relations (\ref{newtest}) between the
$P$, $T$ and $S$ amplitudes hold for neutral decays are
not changed by isospin conserving final state interactions.

Further analysis of the contribution of this additional $I=1/2$ contribution
to final state interactions is needed to obtain
definite values for CP violation in charged $B$ decay.

\section {Comparison with other approaches}

Previous analyses \cite{nurosgro,ROSGRO} were performed at a time when experimental
values for $B\rightarrow K\pi$ branching ratios were not
sufficiently precise to enable a significant test of the sum rule
(\ref{eqapprev}). Values of each of the three interference terms in (\ref{newtest})
were statistically consistent with zero.
The full analysis required the use of data from $B\rightarrow \pi\pi$ decays and
the assumption of $SU(3)_{flavor}$ symmetry. Contributions of the electromagnetic
penguin diagram were included and the relevant CKM matrix elements were included. But
there was no inclusion of constraints from the Pauli principle nor contributions from
final state interactions.

The present analysis uses new experimental data which enable a statistically significant
evaluation of the interference terms (\ref{newtest}) without additional information
from $B\rightarrow \pi\pi$ decays or the assumption of $SU(3)_{flavor}$ symmetry.
Contributions from all isospin invariant finite state
interactions are included as well as constraints from the Pauli principle. The
flavor topology definition of the interference parameters includes contributions from the electromagnetic penguin diagram since the  quark states in final state of a photon can be rewritten as the sum of an isoscalar and
a $u\bar u$ state. However the flavor topology parameters are no longer simply related to the
CKM matrix elements. Additional assumptions and information are necessary to determine the CKM matrix elements
and explain CP violation.

The main advantage of this approach is that it gives a simple explanation for the experimental value
 (\ref{newtest0}) and the vanishing of the experimintal value
\beq{vanish}%
\frac{2B(B^o\rightarrow K^o \pi^o) - B(B^o \rightarrow K^+ \pi^-)} {{{[\tau^o}/{\tau^+]}}\cdot[ B(B^+ \rightarrow K^o \pi^+) + 2B(B^+ \rightarrow K^+ \pi^o)] -
2B(B^o \rightarrow K^+ \pi^- )}$$
=0.09\pm 0.1
\eeq

This vanishing of tree-penguin interference $B^+$ decays is explained by a symmetry analysis including the constraints of the Pauli principle on states containing a pair of identical $u$ quarks.
\section {Conclusion}

Experiment has shown that the penguin-tree interference contribution in $B^+
\rightarrow K^+\pi^o$ decay is very small and may even vanish. The
corresponding interference contributions to neutral $B\rightarrow K\pi$ decays
have been shown experimentally to be finite. In charged decays the
previously neglected Pauli antisymmetrization  produces a cancelation
between color-favored and color-suppressed tree diagrams which differ by the
exchange of identical $u$ quarks. This explains the smallness of penguin-tree interference and
small CP violation in charged $B$ decays. Pauli cancelation does not occur in neutral decay diagrams
which have no pair of identical quarks.
 This can explain why CP violation
has been observed in neutral $B \rightarrow K\pi$ decays and not in charged
decays



{\begin{figure}[htb]
$$\beginpicture
\setcoordinatesystem units <\tdim,\tdim>
\stpltsmbl
\putrule from -25 -30 to 50 -30
\putrule from -25 -30 to -25 30
\putrule from -25 30 to 50 30
\putrule from 50 -30 to 50 30
\plot -25 -20 -50 -20 /
\plot -25 20 -50 20 /
\plot 50 20 120 40 /
\plot 50 -20 120 -40 /
\springru 50 0 *3 /
\plot 120 20 90 0 120 -20 /
\put {$\overline{b}$} [b] at -50 25
\put {${q_s}$} [t] at -50 -25
\put {$\overline{s}$} [l] at 125 40
\put {$\overline{n}$} [l] at 125 -20
\put {$n$} [l] at 125 20
\put {${q_s}$} [l] at 125 -40
\put {$\Biggr\}$ $K(\vec k)$} [l] at 135 30
\put {$\Biggr\}$  $\pi(-\vec k)$} [l] at 135 -30
\put {$G$} [t] at 70 -5
\setshadegrid span <1.5\unitlength>
\hshade -30 -25 50 30 -25 50 /
\linethickness=0pt
\putrule from 0 0 to 0 60
\endpicture$$
\caption{\label{fig-2}} \hfill ``Gluonic penguin'' ($P$) diagram.
$G$ denotes any number of
gluons. $n$ denotes $u$ or $d$ quark.
\hfill~ \end{figure}}

{\begin{figure}[htb]
$$\beginpicture
\setcoordinatesystem units <\tdim,\tdim>
\stpltsmbl
\putrule from -25 -30 to 50 -30
\putrule from -25 -30 to -25 30
\putrule from -25 30 to 50 30
\putrule from 50 -30 to 50 30
\plot -25 -20 -50 -20 /
\plot -25 20 -50 20 /
\plot 50 0 120 -20 /
\plot 50 -20 120 -40 /
\photonru 50 20 *3 /
\plot 120 40 90 20 120 20 /
\put {$\overline{b}$} [b] at -50 25
\put {$q_s$} [t] at -50 -25
\put {$\overline{s}$} [l] at 125 40
\put {$u$} [l] at 125 20
\put {$\overline{u}$} [l] at 125 -20
\put {$q_s$} [l] at 125 -40
\put {$\Biggr\}$  $K(\vec k)$} [l] at 135 30
\put {$\Biggr\}$  $\pi(-\vec k)$}  [l] at 135 -30
\put {$W$} [t] at 70 15
\setshadegrid span <1.5\unitlength>
\hshade -30 -25 50 30 -25 50 /
\linethickness=0pt
\putrule from 0 0 to 0 60
\endpicture$$
\caption{\label{fig-4}} \hfill Color favored tree ($T$) diagram.
 \hfill~ \end{figure}}

{\begin{figure}[htb]
$$\beginpicture
\setcoordinatesystem units <\tdim,\tdim>
\stpltsmbl
\putrule from -25 -30 to 50 -30
\putrule from -25 -30 to -25 30
\putrule from -25 30 to 50 30
\putrule from 50 -30 to 50 30
\plot -25 -20 -50 -20 /
\plot -25 20 -50 20 /
\plot 50 20 120 40 /
\plot 50 -20 120 -40 /
\photonru 50 0 *3 /
\plot 120 20 90 0 120 -20 /
\put {$\overline{b}$} [b] at -50 25
\put {${q_s}$} [t] at -50 -25
\put {$\overline{u}$} [l] at 125 40
\put {$u$} [l] at 125 20
\put {$\overline{s}$} [l] at 125 -20
\put {${q_s}$} [l] at 125 -40
\put {$\Biggr\}$ $\pi(\vec k)$}  [l] at 135 30
\put {$\Biggr\}$ $K(-\vec k)$} [l] at 135 -30
\put {$W$} [t] at 70 -5
\setshadegrid span <1.5\unitlength>
\hshade -30 -25 50 30 -25 50 /
\linethickness=0pt
\putrule from 0 0 to 0 60
\endpicture$$
\caption{\label{fig-5}} \hfill Color suppressed tree ($S$) diagram.
\hfill~ \end{figure}}

\section*{Acknowledgements}

This research was supported in part by the U.S. Department of Energy, Division
of High Energy Physics, Contract DE-AC02-06CH11357. It is a pleasure to thank
Michael Gronau, Yuval Grossman, Marek Karliner, Zoltan Ligeti, Yosef Nir,
Jonathan Rosner, J.G. Smith, and  Frank Wuerthwein for discussions and
comments.

%
\catcode`\@=11 
\def\references{
\ifpreprintsty \vskip 10ex
%
\hbox to\hsize{\hss \large \refname \hss }\else
\vskip 24pt \hrule width\hsize \relax \vskip 1.6cm \fi \list
{\@biblabel {\arabic {enumiv}}}
{\labelwidth \WidestRefLabelThusFar \labelsep 4pt \leftmargin \labelwidth
\advance \leftmargin \labelsep \ifdim \baselinestretch pt>1 pt
\parsep 4pt\relax \else \parsep 0pt\relax \fi \itemsep \parsep \usecounter
{enumiv}\let \p@enumiv \@empty \def \theenumiv {\arabic {enumiv}}}
\let \newblock \relax \sloppy
 \clubpenalty 4000\widowpenalty 4000 \sfcode `\.=1000\relax \ifpreprintsty
\else \small \fi}
\catcode`\@=12 

\end{document}